\documentclass[11pt,english]{article}
\usepackage{amsmath}
\usepackage{amsthm}
\usepackage{mathptmx}
\usepackage{newtxmath}
\usepackage[T1]{fontenc}
\usepackage[latin9]{inputenc}
\usepackage{geometry}
\usepackage{color}
\usepackage{babel}
\usepackage{float}
\usepackage{url}
\usepackage{graphicx}
\usepackage{setspace}
\usepackage[authoryear]{natbib}
\usepackage[unicode=true,pdfusetitle,
 bookmarks=true,bookmarksnumbered=false,bookmarksopen=false,
 breaklinks=false,pdfborder={0 0 0},pdfborderstyle={},backref=false,colorlinks=true]
 {hyperref}
\hypersetup{
 linkcolor=red,urlcolor=red,citecolor=blue}

\usepackage{eurosym}

\makeatletter

\providecommand{\tabularnewline}{\\}

\usepackage{babel}
\usepackage[bottom]{footmisc}
\usepackage{indentfirst}
\usepackage{lscape}
\usepackage{xcolor}
\usepackage{titlesec}
\date{August 2024}

\@ifundefined{showcaptionsetup}{}{%
 \PassOptionsToPackage{caption=false}{subfig}}
\usepackage{subfig}
\makeatother

\begin{document}
\title{Inefficiencies of Carbon Trading Markets\footnote{Nicola Borri, LUISS University, email: nborri@luiss.it. Yukun Liu, University of Rochester, email: yliu229@ur.rochester.edu. Aleh Tsyvinski, Corresponding author, Department of Economics at Yale University, email: a.tsyvinski@yale.edu. Xi Wu, University of California Berkeley, email: xiwu@berkeley.edu}}
\author{Nicola Borri, Yukun Liu, Aleh Tsyvinski, and Xi Wu}

\maketitle
\begin{abstract}
\noindent The European Union Emission Trading System is a prominent market-based mechanism to reduce emissions. While the theory is well-understood, we are the first to study the whole cap-and-trade mechanism as a financial market. Analyzing the universe of transactions in 2005-2020 (more than one million records of granular transaction data), we show that this market features significant inefficiencies undermining its goals. First, about 40\% of firms never trade in a given year. Second, many firms only trade during surrendering months, when compliance is immediate and prices are predictably high. Third, a number of operators engage in speculative trading, exploiting private information. 

\end{abstract}
\thispagestyle{empty}

\newpage

\setcounter{page}{1}

Carbon emissions are a classic example of externalities, where economic agents
do not fully internalize their environmental impact, resulting in overproduction
of carbon emissions. A cap-and-trade system, where
agents can trade allowances to emit carbon, holds significant promise
and offers a potential market-based solution for firms to internalize the environmental
impact~\citep{goulder2013markets,schmalensee2013so2}. A recent meta-analysis of 70 carbon-pricing schemes~\citep{dobbeling2024systematic} concludes that carbon pricing is effective in substantially reducing emissions. 

While effectiveness in reducing emissions is an important criterion of success, an equally important measure of success of a market-based mechanism is its efficiency. The efficiency of the carbon trading market as a financial market is inherently a question that needs to be addressed by finance tools and, more specifically, by asset pricing tools, as carbon is a tradable asset. 

We document that the European Union Emission Trading System (EU ETS)---the first major emission allowance market and a cornerstone of the EU emissions policy---is strikingly inefficient as a market-based mechanism, which leads to a number of unintended consequences that substantially undermine its intended purpose. First, a large fraction of operators do not trade. Second, many operators time their trades inefficiently, incurring significant additional losses. Third, our analysis uncovers the existence of a number of regulated firms acting as speculators that obtain significant profit by exploiting the market with private information. 

We obtain granular transaction and compliance data of companies
participating in the EU ETS from the European Union Transaction Log
(EUTL). Our sample contains the universe of transactions and compliance
information from the EUTL from February 2005 to April 2020, covering the
first three phases of the EU ETS. There are three main groups of participants
in the market: regulated firms, intermediaries, and regulators. They
are largely represented in the EU ETS market by operator holding accounts,
person holding accounts, and administrative accounts, respectively.
Our analyses focus on the regulated firms (or operators) and study their transactions
with other regulated firms and intermediaries. We exclude transactions
with administrative accounts, as these represent the receiving or
surrendering of emissions allowances to regulators, which is not the
focus of this paper. 

First, the analysis of the trading pattern of the emission allowance market reveals that a significant fraction of firms do not engage
in trading each year---approximately 40\% of the overall sample.
One possible justification for the documented large fraction of non-trading firms
is that they receive an amount of free allowances closely
matching their surrender requirements as part of the EU ETS. However, we find this not to be true in the data. In fact, many of these firms exhibit imbalances of allowances in the years they do not trade at all. The large fraction of non-trading firms is not only a sign of market inefficiency but also, to a significant extent, defeats its purpose as in any cap-and-trade model, for the system to efficiently allocate allowances to the most productive use, firms must participate in the market and actively trade.

Second, we demonstrate that many firms trade at inefficient times and at inefficient prices, leading to additional significant losses. The EU ETS market operates on a fixed verifying and surrendering schedule each year. By the end of April, firms must surrender a sufficient
number of emission allowances to the administrator account to match
their verified emissions. This fixed schedule leads to a predictable pattern
in emission allowance trading, with April consistently showing the
largest net purchases of allowances by regulated firms throughout
the sample period. The magnitude of these purchases is striking, with
net purchases in April being an order of magnitude larger than in
typical non-April months. This high demand for emission allowances
in the surrendering month results in significant price increases,
averaging more than 10\% in Aprils. This empirical evidence reveals that, in addition to a large fraction of non-trading firms, many firms that do trade primarily engage in trading exclusively during the surrendering month, when compliance is immediate, and the price of emission allowances is predictably high. A back-of-the-envelope calculation reveals that this surrendering pattern alone leads to an implied loss for the regulated firms of about \euro5 billion. 

Third, we find that about 10\% of the firms trade an excessive
amount of emission allowances each year, and show that these firms possess private information and are able to exploit it for profit. Using a conservative definition,
these firms have trading volumes more than four times the amount they
surrender in a given year. We provide evidence that the excessive trading is related to these firms exploiting their private information about the market. In fact, these firms purchase emission allowances correctly, anticipating prices increase in the future, and vice versa, sell emission allowances correctly, anticipating future price decreases. To capture the net purchases of firms
in a month, we construct a variable called ``$Speculator \ Ratio$'', defined
as the net flow from person holding accounts to operator holding accounts
divided by the total transaction between the two types of accounts
for each month. We establish that $Speculator \ Ratio$ is a strong predictor
of future cumulative emission allowance returns. Using $Speculator \ Ratio$ to predict cumulative allowance returns from one month to twelve
months ahead, we find that the coefficient estimates are all positive,
becoming significant at the 5\% level at the two-month horizon. The results are consistent with the notion that these firms possess private information and tend to accumulate emission allowances when they anticipate future price increases. A back-of-the-envelope estimation reveals that these firms make about \euro8 billion in profit during the sample period.

If the observed excessive trading by some firms is driven by asymmetric
information and operators exploiting private information for speculative trading,
we would expect the return predictability power to be stronger for
trades conducted by firms that more frequently transact in the emission
allowance market. Our analysis reveals that
the return predictability is entirely driven by the transactions conducted
by firms that frequently transact in the emission allowance market. This evidence further supports the notion that these firms possess private information
and are able to exploit it for profit.

\section{Market Overview and Data\label{sec:Data}}

This section provides an overview of the EU ETS market and describes
the comprehensive data collected on the universe of transactions within
the EU ETS market.

\subsection{Market Overview}

The EU ETS is the first large carbon market in the world. The system
covers more than 13,000 facilities spanning diverse industrial sectors.
Functioning as a cap-and-trade system, the EU ETS allocates emission
permits, commonly known as European Union Allowances (EUAs), to participating
companies. Participating companies are required to report their annual
carbon emission amount to the regulatory authorities and also must
surrender a corresponding amount of permits. 

The EU ETS was launched in 2005 as a cornerstone of the EU policy
to reduce greenhouse gas emissions. The evolution of the EU ETS is
organized in various ``phases''. The first trading phase (Phase
I) goes from 2005 to 2007. This is a pilot phase
that lays the groundwork for subsequent phases by establishing trading
rules and procedures. The second trading phase (Phase II) goes from 2008 to 2012. Building upon Phase I, the EU ETS expanded
its scope and refined its mechanisms, including efforts to standardize
the allocation rules and improve the pricing and functionality of
the trading system. The third phase (Phase III) goes
from 2013 to 2020, while the current phase (Phase IV) started in 2021
and will end in 2030.

During our sample period, the EU ETS has a fixed verifying
and surrendering schedule each year. In particular, each year operators
must submit an emission report. The emissions data for a given year
must be verified by an accredited verifier by the end of March of
each year. Then, operators must surrender the equivalent number of
allowances to match the installation's emissions by the end of April.
In other words, April is the month of surrendering for all firms under the EU ETS. For each ton of emissions where
no allowance is surrendered on time, there is a penalty of \euro100,
in addition to the cost of surrendering allowances. The names of penalized
operators are also made public.

\subsection{Data}

\subsubsection{EU ETS Trading and Compliance Data}

The EU ETS individual trading and compliance data are publicly available
and are from the European Union Transaction Log (EUTL)~\citep{abrell_database}.\footnote{The data used in the paper is publicly available through the EUTL website (\href{https://ec.europa.eu/clima/ets/}{https://ec.europa.eu/clima/ets/}). We downloaded the data using the tools available at \href{https://www.euets.info}{https://www.euets.info}. The EUA spot price is from Datastream.} The EUTL serves
as the primary platform for reporting and monitoring within the EU
ETS. This tool enables the European Commission to transparently disclose
crucial data regarding compliance by regulated entities, the participation
of stakeholders within the system, and the transactions conducted
among these participants. Our sample contains the universe of transactions
and compliance information from the EUTL from February 2005 to April 2020,
spanning the first three phases of the EU ETS. The EUTL data that
we use in this paper contain a number of components, including compliance,
account, and transaction information, which we describe separately
below.

\paragraph*{Compliance Information}

Compliance within the EU ETS occurs at the level of installations. The installations have the obligation to surrender emission allowances
for the verified amounts. The EUTL data contain detailed information
about the identities of the installations as well as the compliance
information. The compliance information contains the allocated amount
and the surrendered amount for each installation in each year. In
the sample, the compliance rate of installations is close to 100\%,
and therefore, non-compliance is unlikely to be an issue in our analysis. 

\paragraph*{Account Information}

There are three basic account types in the EUTL
~\citep{abrell_database}: operator
holding accounts, person holding accounts, and administrative accounts.
An operator holding account represents an installation---the regulated
participant. The operator holding account allows the installation
to receive, transfer, and surrender emission allowances. Non-regulated
participants (or intermediaries) can participate in the EU ETS by using personal holding
accounts to trade emission allowances. Regulators must use administrative
accounts to allocate and receive surrendered emission allowances. Accounts are held by account holders or companies. Sometimes, a company may have multiple accounts, including operator holding accounts and/or person holding accounts.  

\paragraph*{Transaction Information}

The EUTL records all emission allowance transactions between accounts.
Transactions can happen between any two accounts, including operator,
person, and administrative accounts. Reallocations of emission allowances
within the same firm are recorded as transactions in the EUTL. Receiving
and surrendering emission allowances from and to administrative accounts
are also recorded as transactions. In this paper, we exclude any EUTL recorded transaction that involves administrative accounts, as our focus is on the actual transactions, especially those involving regulated entities, instead of the allocated or surrendering behaviors in the EU ETS. Furthermore, since we aggregate transaction information at the company level, transactions recorded
within the same company in the EUTL always net out to zero. 

\begin{figure}[H]
\caption{EUA Buy and Sell Transactions by Holding Type\label{fig:EUA-Buy-and-Sell}}

\medskip{}

This figure presents the evolution in the total amount (top panels)
and transaction counts (bottom panel) of EUA sold and purchased by
holding type. Data is from February 2005 to April 2020.

\medskip{}

\centering{}\includegraphics[scale=0.5]{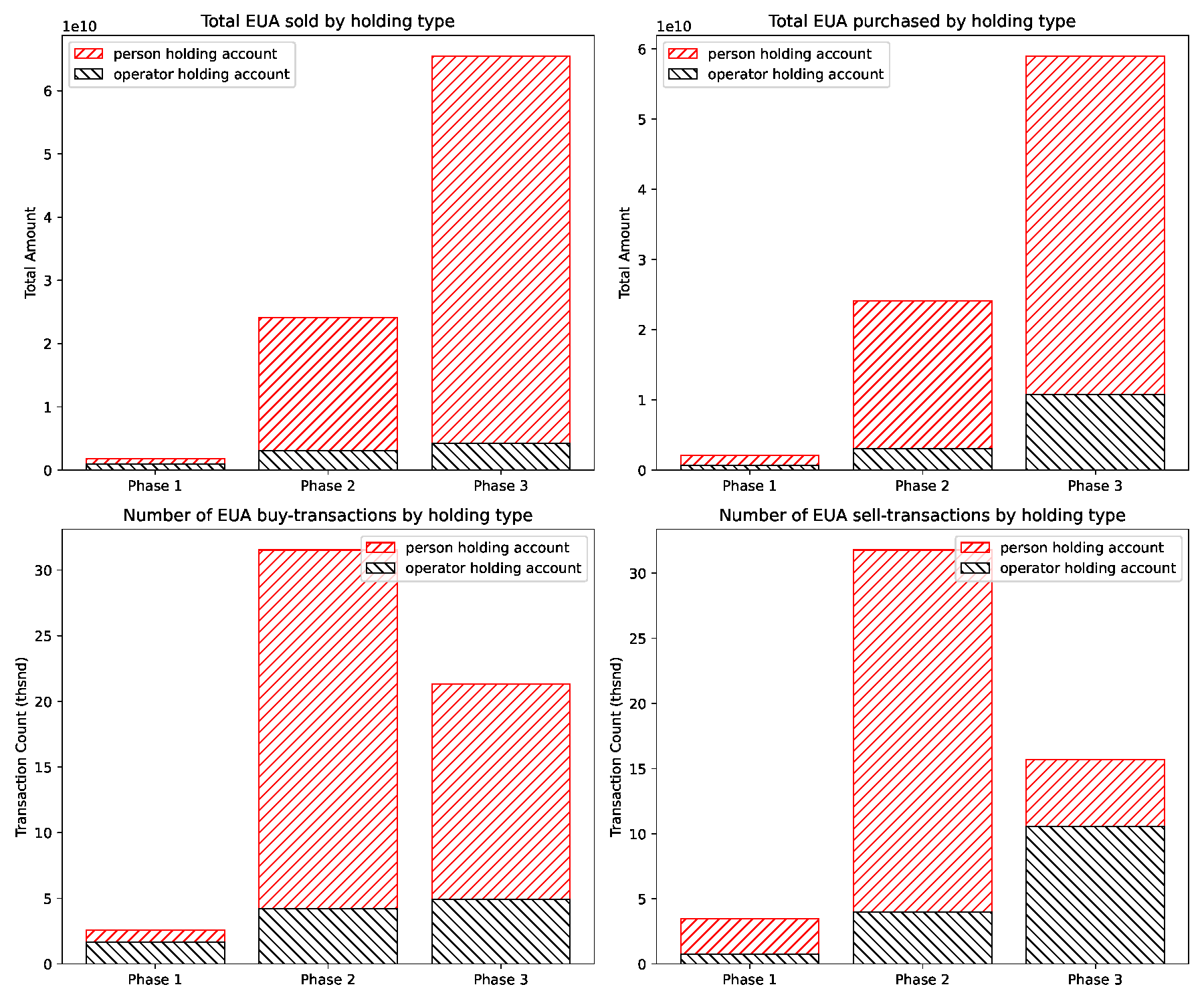}
\end{figure}

The number of person holding accounts was at
its lowest (334 accounts) at the onset of EU ETS. This number increased,
peaking at over 1,500 at the beginning of Phase II, and then steadily
declined almost every year, reaching 695 in 2020. In contrast, the
number of operator holding accounts has been increasing, starting
at 904 at the onset of the EU ETS, reaching a peak of 6,033 in 2013,
and stabilizing around 4,000 in 2020. While the number of person holding
accounts is a fraction of the number of operator holding accounts,
person holding accounts represent the largest share of the total volume
of EUAs sold and purchased, as illustrated in Figure \ref{fig:EUA-Buy-and-Sell}.
Furthermore, the figure shows that the total amount of EUAs sold and
purchased by person holding accounts has been increasing from Phase
I to Phase III of the EU ETS, reaching approximately 65 billion
EUAs sold (\euro683 billion) and 58 billion EUAs purchased (\euro612 billion) in Phase III. Operator
holding accounts, instead, account for approximately 4 billion
in EUAs sold (\euro41 billion) and 10 billion in EUAs purchased (\euro111 billion) in Phase III.\footnote{We use the daily EUA spot price, obtained from Datastream, to estimate the euro value of the EUAs purchased and sold.}

Table \ref{tab:Descriptive-Statistics-of-Operators-Trading} provides
descriptive statistics on operator trading throughout the entire sample
period, as well as across the three distinct phases of the EU ETS
market. Panel A shows that, on average, an operator sold approximately
624 thousand EUAs and purchased 1,150 thousand EUAs over the entire
sample. The average share of EUAs sold to other operators was 22\%,
while the share of EUAs purchased from other operators was 17\%. Overall,
the total trading activity (combining both sales and purchases) of
the average operator amounted to around 1,774 thousand EUAs, with
18\% of this total involving transactions with other operators. The
proportions of sales, purchases, and total trade involving other operators
are quite consistent across the three market phases. Panel B highlights
that the distribution of sales and purchases by operators is right-skewed,
meaning that the median quantities are substantially smaller. Specifically,
the median operator sold or bought about 20 EUAs and interacted primarily
with intermediaries rather than with other operators. This pattern
remains consistent across the different phases of the market.

\begin{table}[H]
\caption{Operator Trading in the EU ETS Market\label{tab:Descriptive-Statistics-of-Operators-Trading}}
{\footnotesize{}\medskip{}
}{\footnotesize\par}

This table shows the fraction of trades operators engage with other
operators. Panel A reports the average numbers of sales, buys, and
total trades as well as the fractions of sales, buys, and total trades
with other operators. Panel B reports the median numbers of sales,
buys, and total trades as well as the fractions of sales, buys, and
total trades with other operators. 

{\footnotesize{}\medskip{}
}{\footnotesize\par}
\centering{}%
\begin{tabular}{ccccccc}
\hline 
 & Sales & To Operators (\%) & Buys & From Operators (\%) & Total Trades & With Operators (\%)\tabularnewline
\hline 
 & \multicolumn{6}{c}{Panel A: Averages}\tabularnewline
\hline 
 &  &  &  &  &  & \tabularnewline
Overall & 624.69 & 0.22 & 1150.28 & 0.17 & 1774.97 & 0.18\tabularnewline
Phase 1 & 1455.94 & 0.18 & 1863.16 & 0.22 & 3319.1 & 0.19\tabularnewline
Phase 2 & 991.99 & 0.18 & 1815.87 & 0.21 & 2807.86 & 0.2\tabularnewline
Phase 3 & 507.11 & 0.25 & 989.58 & 0.16 & 1496.69 & 0.17\tabularnewline
 &  &  &  &  &  & \tabularnewline
\hline 
 & Sales & To Operators (\%) & Buys & From Operators (\%) & Total Trades & With Operators (\%)\tabularnewline
\hline 
 & \multicolumn{6}{c}{Panel B: Medians}\tabularnewline
\hline 
 &  &  &  &  &  & \tabularnewline
Overall & 20.36 & 0 & 19.5 & 0 & 57.32 & 0\tabularnewline
Phase 1 & 48.46 & 0 & 0 & 0.01 & 95 & 0\tabularnewline
Phase 2 & 50.93 & 0 & 29.74 & 0 & 110.24 & 0.01\tabularnewline
Phase 3 & 10.16 & 0 & 24 & 0 & 60.95 & 0\tabularnewline
 &  &  &  &  &  & \tabularnewline
\hline 
\end{tabular}
\end{table}

\section{Trading Patterns\label{sec:Trading-Patterns}}

In this section and the next, we investigate the trading patterns of firms in the EU ETS. Our analysis documents that there is significant inefficiency in the emission allowance trading market.

\subsection*{Trading patterns of the average firm}

\begin{figure}[H]
\caption{Distribution of Operator Trading Amounts\label{fig:trading-frequency}}

\medskip{}

The top panels plot the fraction of regulated companies with different trading
amounts in each year, excluding transactions involving administration
accounts. Panels A, B, C, and D show the plots for the overall sample,
the Phase I sample, the Phase II sample, and the Phase III sample,
respectively. For each figure, we group the operators into the following
categories: no trade; 0 $\leq$ trades $<$ 20k; 20k $\leq$ trades
$<$ 40k; 40k $\leq$ trades $<$ 60k; 60k $\leq$ trades $<$ 80k;
and 80k $\leq$ trades. The bottom panels plot the fraction of regulated companies with different trading ratios in each year, where the trading ratio is defined as the amount
traded, excluding trades with administration accounts divided by
its surrendering amount. Panels E, F, G, and H show the plots for
the overall sample, the Phase I sample, the Phase II sample, and the
Phase III sample, respectively. For each figure, we group the operators
into the following categories: no trade; trade ratios < 1; 1 $\leq$
trade ratios $<$ 2; 2 $\leq$ trade ratios $<$ 3; 3 $\leq$ trade
ratios $<$ 4; 4 $\leq$ trade ratios. Data is for the period from
February 2005 to April 2020.

\medskip{}

\subfloat[Overall]{\begin{centering}
\includegraphics[width=4cm]{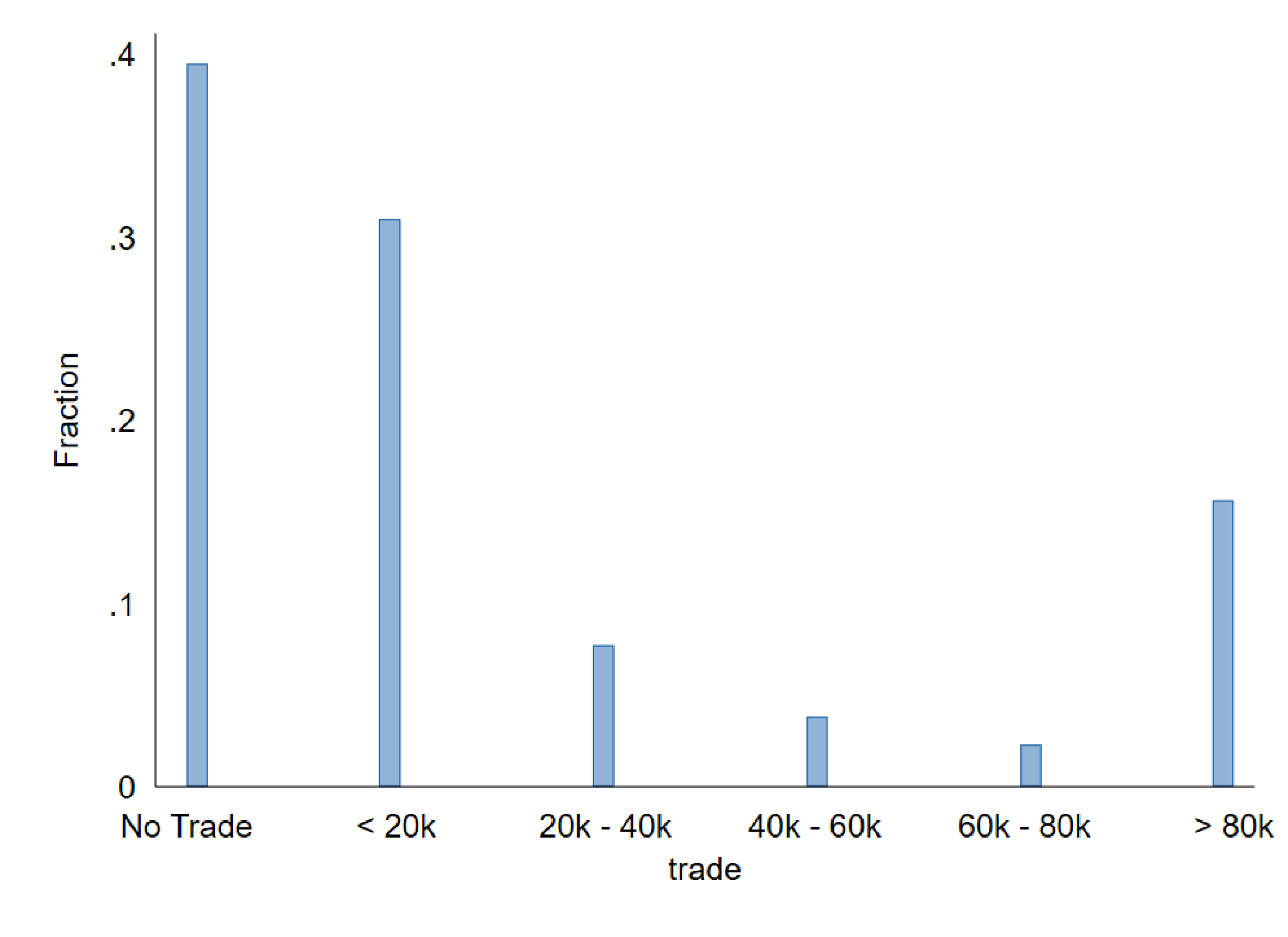}
\par\end{centering}
}
\subfloat[Phase I]{\begin{centering}\includegraphics[width=4cm]{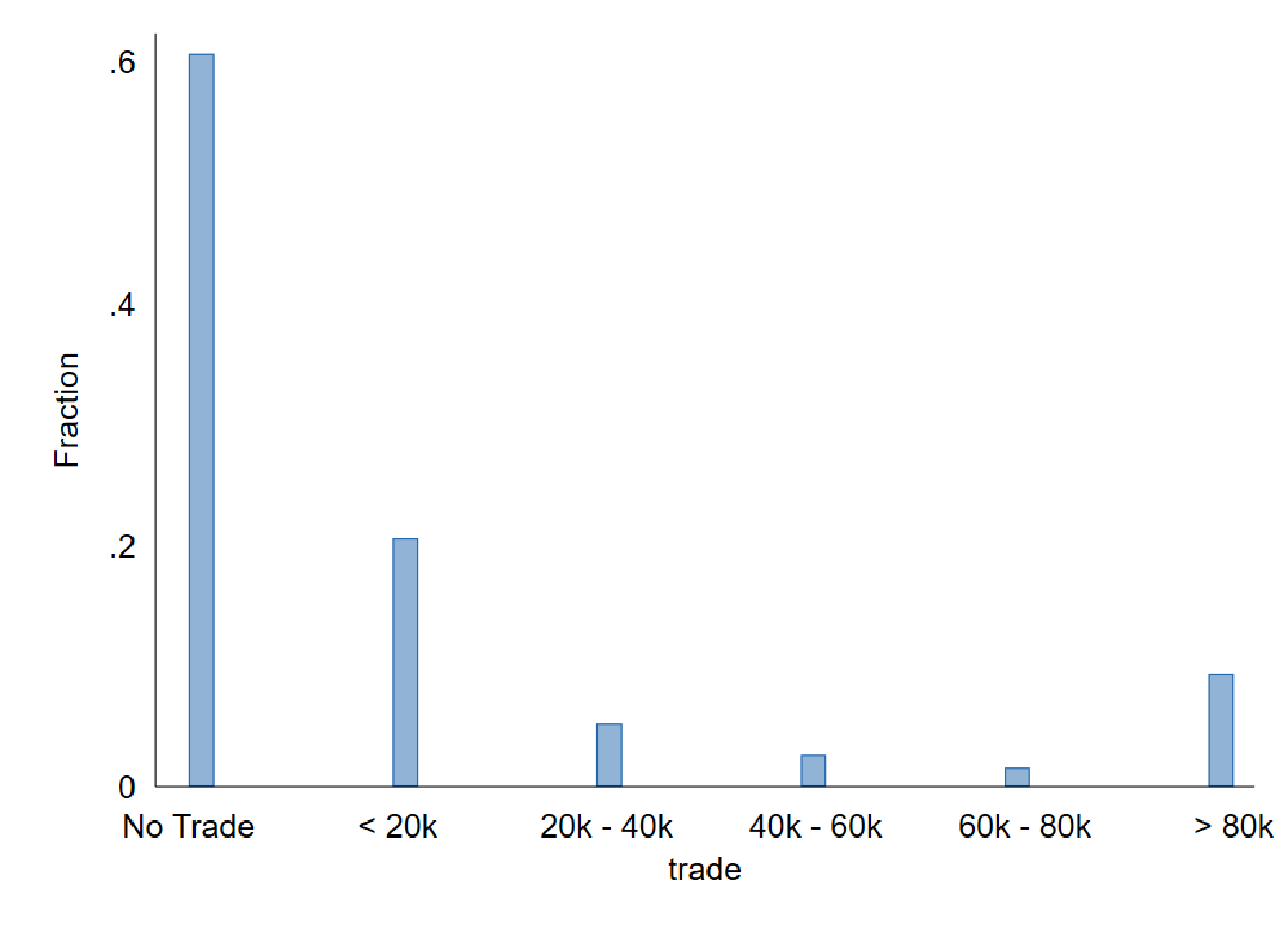}
\par\end{centering}
}
\subfloat[Phase II]{\begin{centering}
\includegraphics[width=4cm]{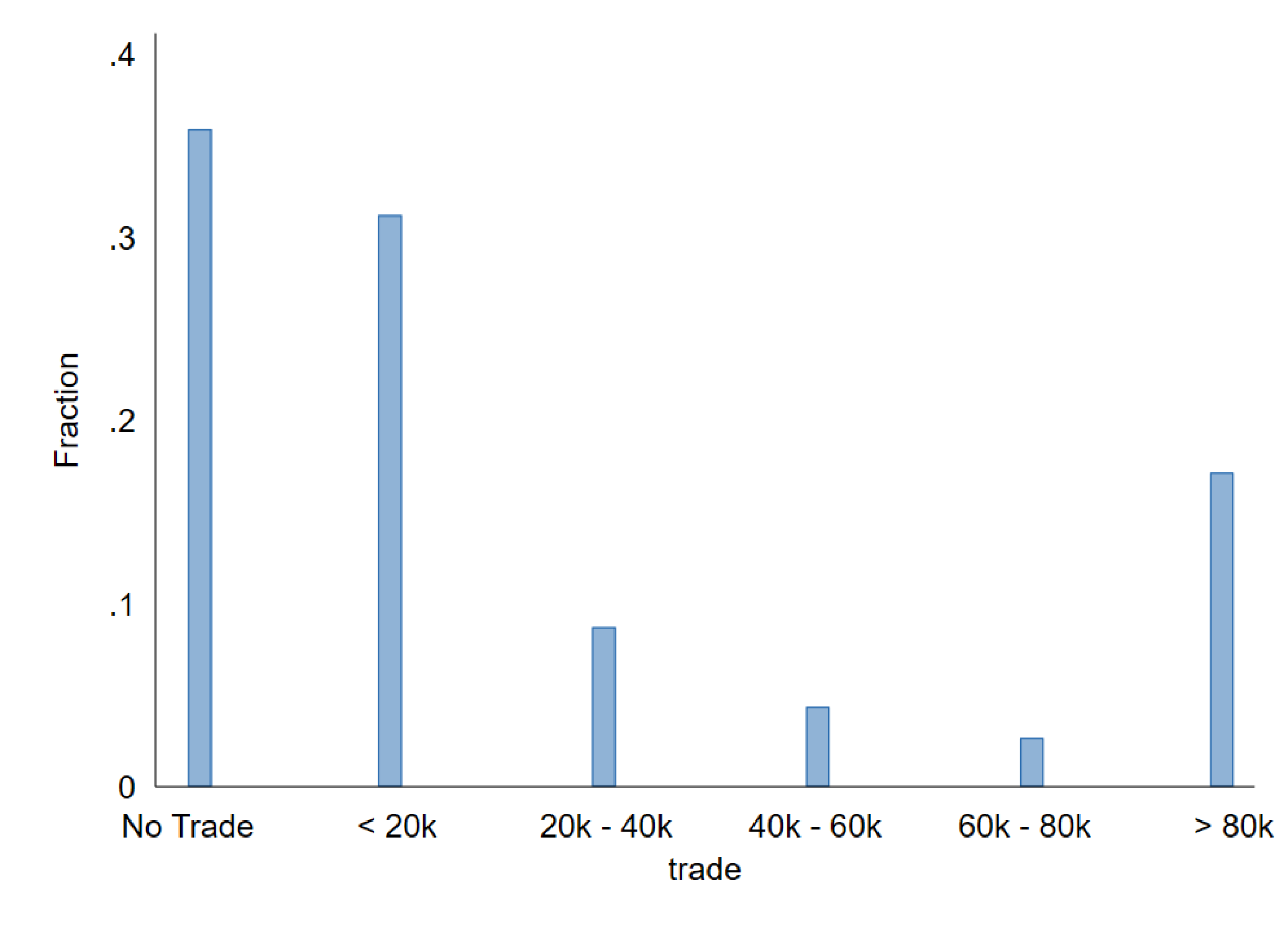}
\par\end{centering}
}
\subfloat[Phase III]{\begin{centering}
\includegraphics[width=4cm]{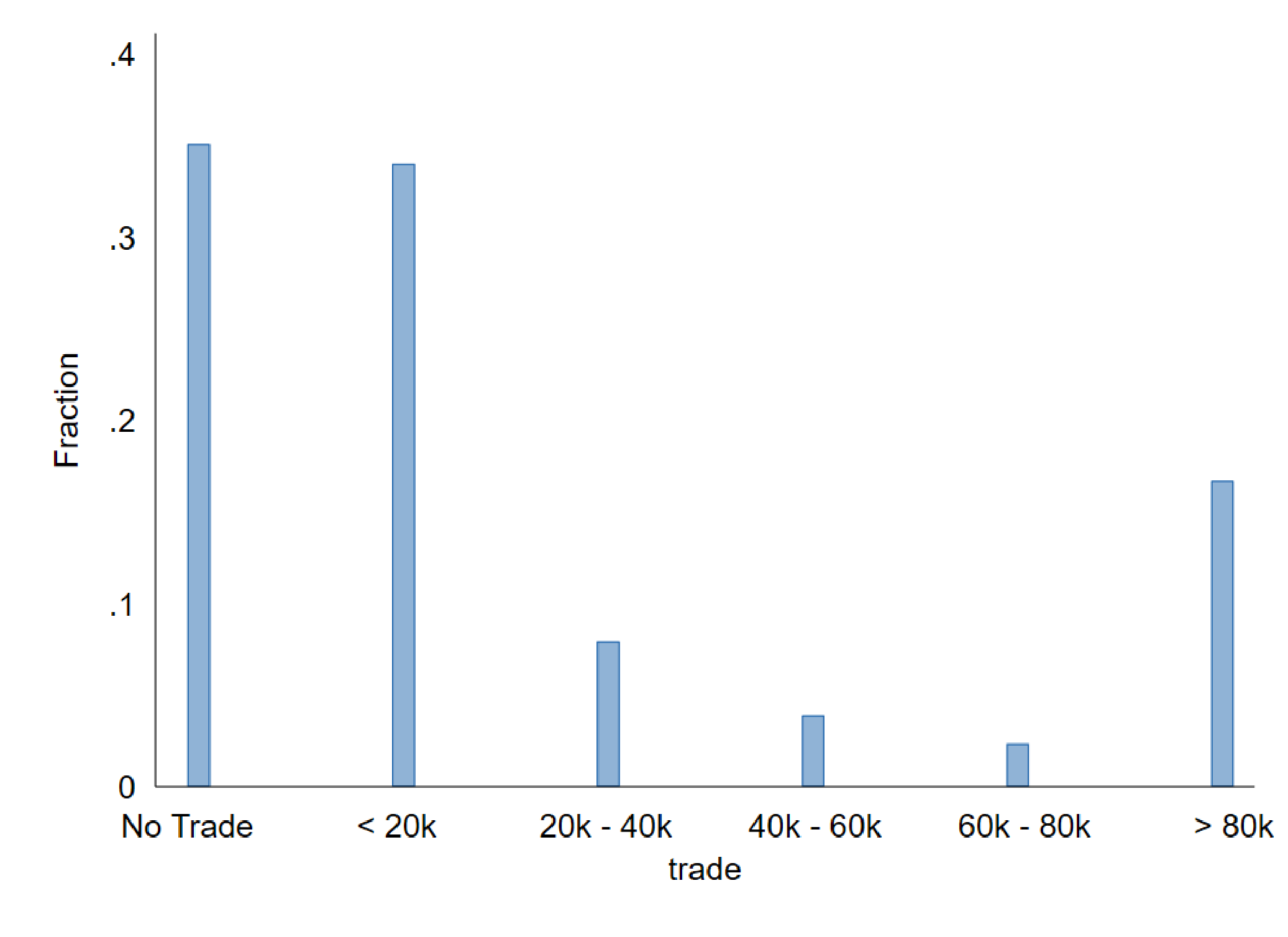}
\par\end{centering}
}

\subfloat[Overall]{\begin{centering}
\includegraphics[width=4cm]{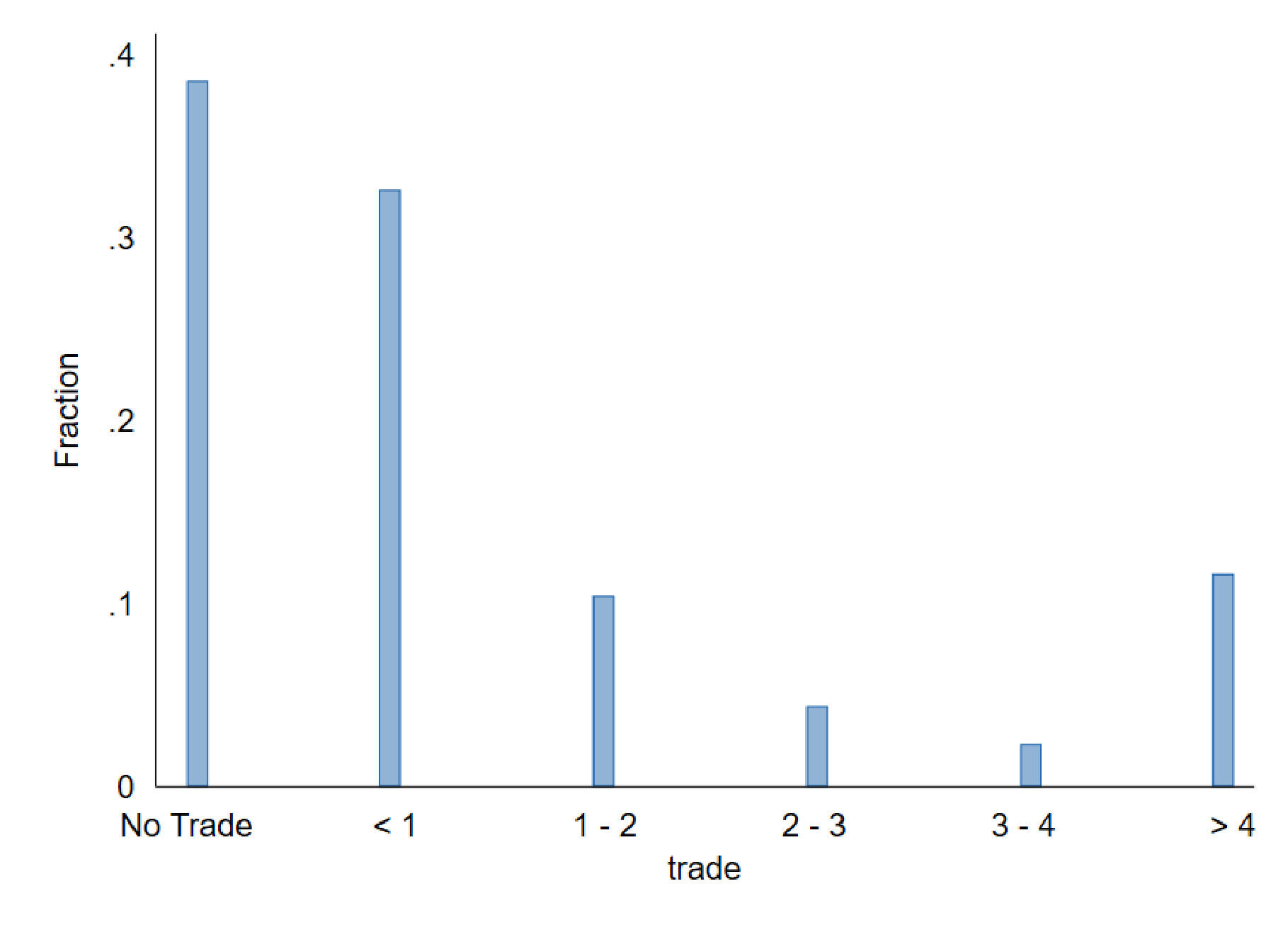}
\par\end{centering}
}
\subfloat[Phase I]{\begin{centering}
\includegraphics[width=4cm]{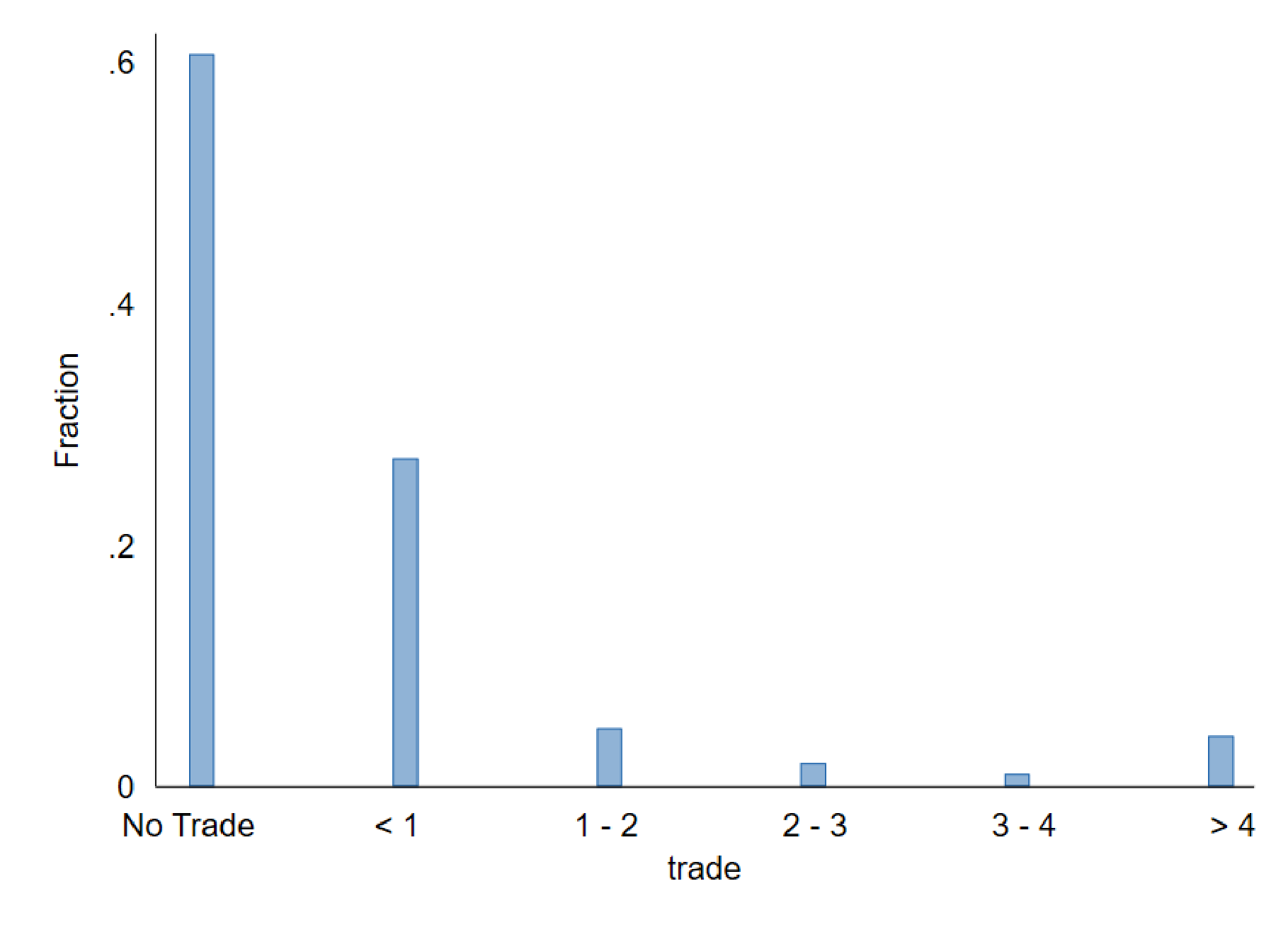}
\par\end{centering}
}
\subfloat[Phase II]{\begin{centering}
\includegraphics[width=4cm]{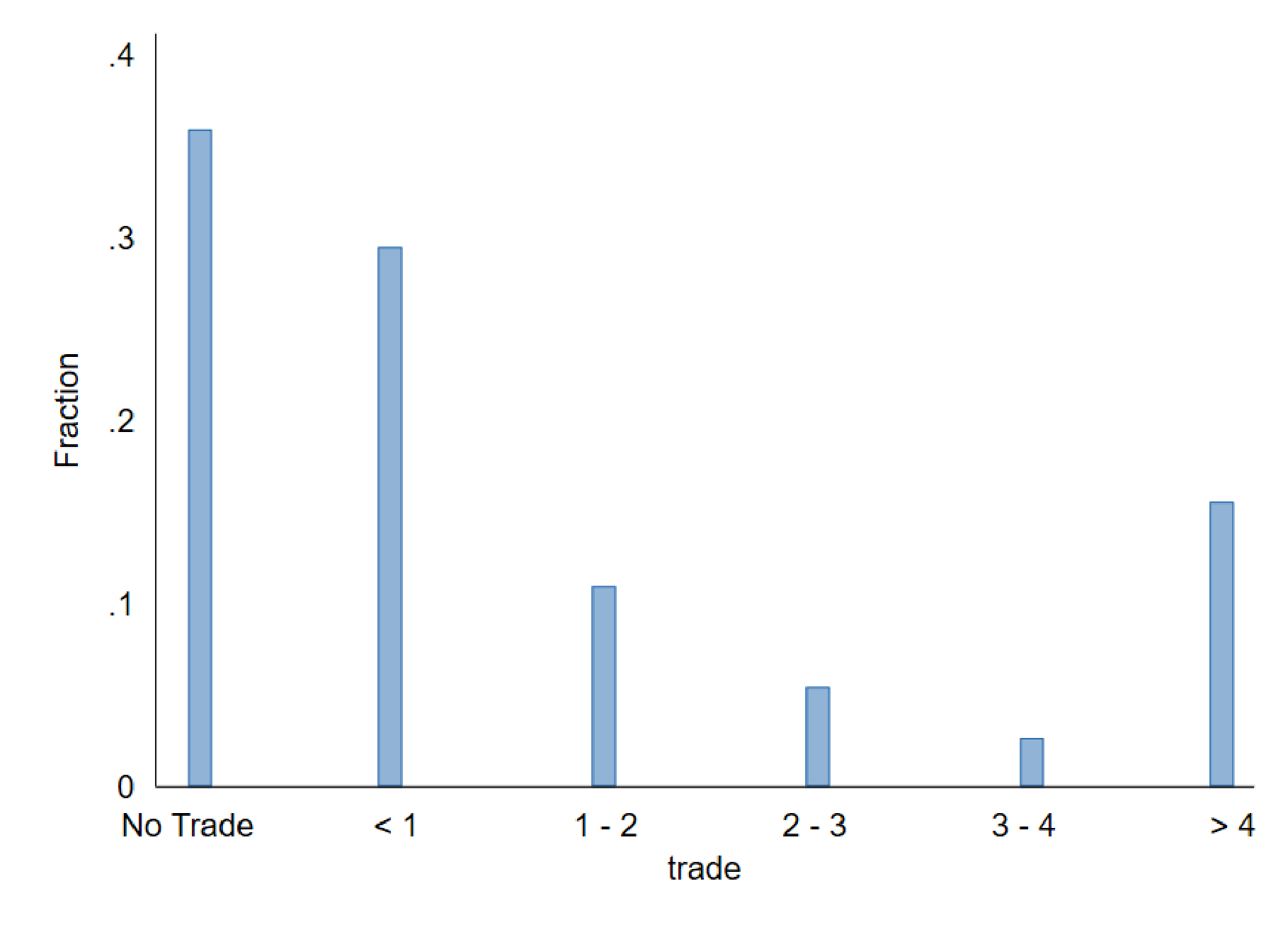}
\par\end{centering}
}\subfloat[Phase III]{\begin{centering}
\includegraphics[width=4cm]{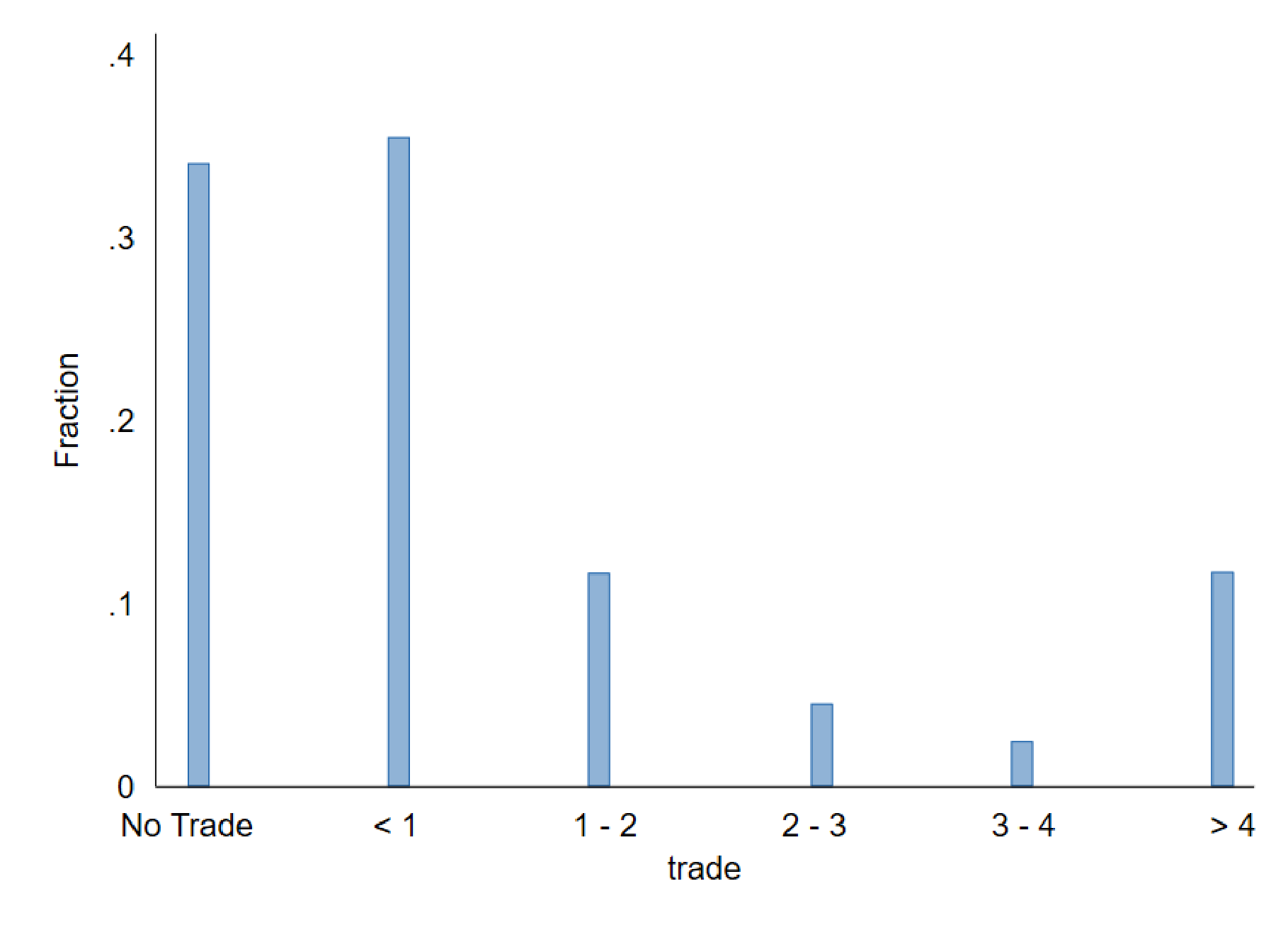}
\par\end{centering}
}
\end{figure}

In any cap-and-trade model, for the system to efficiently allocate allowance to the most productive use, firms must participate in the market and actively trade. We study the trading patterns of firms with operator holding accounts
that participate in the EU ETS. The top panels of Figure \ref{fig:trading-frequency} plot the fraction of operators
with different trading amounts in each year. Panel A refers to the full sample. Panels B to C plot the results for the
samples corresponding to Phase I, Phase II, and Phase III, respectively.
From the figure, we observe that the fraction of firms with no trade
in a year is significant. In the overall sample, this fraction is about 40\%. The fraction is highest in Phase I, approaching 60\%. Even in Phases II
and III, the fraction of operators with no trades in a year exceeds
35\%. The large fraction of non-trading firms significantly undermines the purpose of the cap-and-trade system to minimize costs in achieving pollution reduction by trading emission allowances.

One possible justification for the large fraction of non-trading firms
is that they receive a free amount of allowances similar to the amount they need to surrender. However, the data does not support this explanation. For these firms, the
5th and 25th percentiles of the allocated-to-surrendered allowances ratio are
0 and 0.75, respectively, while the 75th and 95th percentiles are
1.37 and 5.41. This indicates that many of these firms experience
significant imbalances in the years they do not trade.

The average annual trading volume for firms with non-zero trading
is approximately 18 thousand contracts. The total trading volume for
operators exhibits a heavily right-skewed distribution. Among firms
with non-zero trading activity, approximately 30\% trade less than
20 thousand contracts per year, while about 15\% of firms trade more
than 80 thousand contracts annually.

Some operators may trade a large volume of emission contracts simply due to the size of their firm. To address this possibility,
given that larger firms typically surrender more permits, we normalize
the volume a firm trades annually by its surrendering amount. This
is done by calculating the trading ratio, which is the firm's total
trading volume (excluding trades with administration accounts) divided
by its surrendering amount for the year. The results are reported
in the bottom panel of Figure \ref{fig:trading-frequency}. Panel A refers to
the full sample. Panels B to C plot the results for the samples of
Phase I, Phase II, and Phase III, respectively.

From the bottom panel of Figure \ref{fig:trading-frequency}, we find that about 30\% of the firms have a trading ratio greater than zero but less than one. A
significant portion of firms engage in trading volumes that exceed
their surrendering amount for the year, resulting in a trading ratio
above one. In the full sample, more than 10\% of firms trade more
than four times the amount they surrender in a year. Examining the
various phases of the EU ETS, we observe a similar pattern as seen
in the full sample.

The trading patterns reveal two important features of the EU ETS emission
trading market. Firstly, a significant proportion, around 30\%--40\%
of regulated firms, do not participate in any trading within a given
year. These companies with no emission allowance trading have substantially
different initial allocated amounts and emission abatement technologies,
reflecting their diverse industrial backgrounds. Secondly, a substantial
fraction of firms are involved in trading activities that are disproportionate
to their surrendering amount.

\subsection*{Trading in the Surrendering Month}

In this subsection, we explore the emission trading patterns induced
by the regulatory features of the EU ETS. The EU ETS system has a
fixed verifying and surrendering schedule each year. Specifically,
by the end of April, firms must surrender a sufficient number of emission
allowances to the administrator account to match their emissions.
Consequently, April is the designated month for surrendering allowances
for all firms under the EU ETS system.

\begin{figure}[H]
\caption{Trading and Prices in the Surrendering Month\label{fig:Trading-Concentrate}}

\medskip{}

This figure plots the monthly net purchase of emission allowances
by operators and the average emission allowance returns for each month.
Panel A depicts the average total amount of transactions over time,
while Panel B shows the average emission contract returns by month since 2008. Transactions involving admin accounts are excluded. At least one party
in a transaction must be an operator for the transaction to be included
in the analysis.

\medskip{}

\subfloat[Net Transaction Amount]{\begin{centering}
\includegraphics[width=8cm]{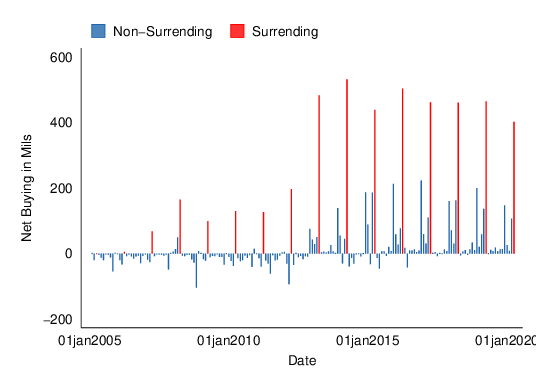}
\par\end{centering}
}
\subfloat[Average EUA Return]{\begin{centering}
\includegraphics[width=8cm]{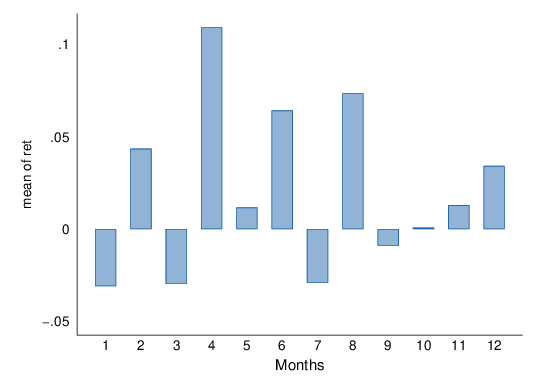}
\par\end{centering}
}
\end{figure}

We compute the monthly net purchase of emission allowances by firms,
excluding transactions involving admin accounts. The results are presented
in Panel A of Figure \ref{fig:Trading-Concentrate}. A striking pattern
emerges: the purchasing of emission allowances predominantly occurs
in April each year, as highlighted in red in the figure. April consistently
shows the largest net purchases of emission allowances by firms throughout
the entire sample period. The difference in magnitudes between surrendering
and non-surrendering months is substantial. The second largest net
purchase month is typically December, possibly due to tax reasons
or the expiration of the most common futures contract on the EUA.

Next, we examine the average monthly emission allowance return, which
we document in Panel B of Figure \ref{fig:Trading-Concentrate}. The
emission allowance return is computed as the monthly log change in
emission allowance prices. We observe that the average return is highest
in April, exceeding 10\%. This monthly return is significantly higher
than returns in other months of the year. These results suggest that
the substantial demand during the surrendering month (April)
contributes to a surge in emission allowance prices. 

To quantify the losses of the regulated firms that purchase emission allowances when prices are predictably high during the surrendering months, we use a back-of-the-envelope calculation of the total implied loss for the regulated firms. Using the total net amount purchased by the regulated firms during the surrendering months, the emission allowance prices at the time, and assuming the average 10\% predictable appreciation for the surrendering months, we estimate that the total loss by the regulated firms amounts to about \euro5 billion, or 2\% of the total traded volume of the regulated firms. 

\section{Emission Allowance Return Predictability\label{sec:Predictability}}

The evidence presented in the previous sections reveals that a significant
portion of firms do not trade in the emission allowance market. Among
those that do trade, many transactions occur only in the surrendering
month, leading to an aggregate price surge during that period. 

In this section, we shift our focus to another key trading
pattern revealed earlier. A significant fraction of firms engage in
a disproportionate amount of emission allowance trading. One
potential reason for this disproportionate trading is that these regulated operators, active in the EU ETS, might possess private information
about the market, which they exploit to generate profits through
speculative trading. We would then expect to observe that emission allowance
prices increase on average following positive net purchases by these
firms, and decrease following negative net purchases, reflecting their
impact on market dynamics.

We observe that the pattern, where a small fraction of firms engage in a disproportionate amount of emission allowance trading, as discussed in Section \ref{sec:Trading-Patterns}, cannot be explained by excessively high (or low) expected future emission allowance prices. If this were the case, all firms would either buy or sell large quantities of emission allowances in the current period, which would not account for the significant heterogeneity in the number of transactions we observed in the data.

\subsection*{Baselines Results}

To capture the net purchases of firms in a month, we construct a variable
called ``$Speculator \ Ratio$'' defined as the net flow from person holding
accounts to operator holding accounts divided by the total transaction
between person holding accounts and operator holding accounts for
each month. We then use $Speculator \ Ratio$ to predict future cumulative monthly
emission allowance returns.

Panel A of Table \ref{tab:Predictability} reports first the baseline results
of predicting cumulative future emission allowance returns from one
month ahead to twelve months ahead using $Speculator \ Ratio$. To account for
biases arising from overlapping returns and autocorrelation, the standard
errors are Newey-West~\citep{newey1987simple} adjusted with $n$ lags where
$n$ is the number of overlapping months. The results indicate that
$Speculator \ Ratio$ positively predicts future cumulative emission allowance
returns. The coefficient estimates are all positive and tend to increase
with the horizon. They become significant at the 5\% level at the
two-month horizon and remain significant at least at the 5\% level
for longer horizons. The R-squared values also tend to increase over
longer horizons, reaching 12.7\% at the twelve-month horizon. These
results are consistent with the notion that firms possess private
information and tend to accumulate emission allowances when they anticipate
future price increases in the market. In Section \ref{subsec:Frequent-vs-Infrequent},
we further show that the predictability is fully driven by firms that
trade more frequently.

From our analysis of trading patterns over time, we observe
that there is a strong seasonality in the firm purchases of emission
allowances. We control
for month fixed effects in the return predictability regressions to
account for any potential seasonality effect influencing the results.
We find stronger results when controlling for month fixed effects.
The coefficient estimates are all positive and significant at the
5\% level or beyond for all horizons, including the one-month-ahead
horizon. The magnitudes of the coefficient estimates tend to be larger
compared to the baseline results.

We quantify the total profit these firms make by exploiting private information. Using the net amount purchased and the ``$Speculator \ Ratio$'' for each period, the emission allowance prices at the time, the estimated coefficient estimates from Table \ref{tab:Predictability}, and assuming a conservative holding period of three months, we estimate a back-of-the-envelope total profit by these firms of about \euro8 billion during the full sample, or about 3.5\% of the traded volume of the regulated firms.

\subsection*{Robustness}
We further provide several robustness checks and confirm the baseline return predictability results. In the return predictability literature, several econometric issues have been documented in long-horizon predictive regressions with overlapping
cumulative returns and persistent regressors. First, 
such regressions can lead to inference problems, particularly
with the $t$-statistics, which may diverge over long horizons~\citep{valkanov2003long}. Second, an inference problem
can arise from using persistent regressors when the shocks to the
dependent and independent variables are correlated~\citep{stambaugh1999predictive}. 

To address the first critique, we employ the $t/\sqrt{T}$
test~\citep{valkanov2003long}. We compute the $t/\sqrt{T}$ test and
simulate the critical values with parameters calibrated to our data~\citep{valkanov2003long}. We show that the coefficient estimates are significant at the 5\% level starting from the two-month-ahead horizon with the $t/\sqrt{T}$ test. To address the second critique, we examine the extent of this bias~\citep{stambaugh1999predictive}
and determine that its economic impact is minor; the bias amounts
to less than 1\% of the coefficient estimate. This is due to the relatively low autocorrelation of the independent variable.

{\footnotesize{}}
\begin{table}[H]
{\footnotesize{}\caption{Time-Series Return Predictability\label{tab:Predictability}}
}{\footnotesize\par}

{\footnotesize{}\medskip{}
}{\footnotesize\par}

This table reports the results of time-series EUA return predictability
tests. In Panel A ``$Speculator \ Ratio$'' is defined as the net flow from person holding
accounts to operator holding accounts divided by the total
transaction from person holding accounts to operator holding accounts
for each month. In Panel B ``$Speculator \ Ratio^{low}$'' and ``$Speculator \ Ratio^{high}$'' are defined as the net flow from person
holding accounts to operator holding accounts divided by the total transactions for each week for the above-median and below-median number of transaction operators, respectively. 
Standard errors are Newey-West~\citep{newey1987simple} adjusted with $n$ lags
where $n$ is the number of overlapping periods. $t$-statistics are
reported in parentheses. {*}, {*}{*}, and {*}{*}{*} denote significance
at the 10\%, 5\%, and 1\% levels, respectively.

{\footnotesize{}\medskip{}
}{\footnotesize\par}
\centering{}%
\begin{tabular}{lccccccc}
\hline  
\multicolumn{8}{c}{Panel A: All Traders}\tabularnewline
& $r_{t\to t+1}$ & $r_{t\to t+2}$ & $r_{t\to t+3}$ & $r_{t\to t+4}$ & $r_{t\to t+6}$ & $r_{t\to t+8}$ & $r_{t\to t+12}$\tabularnewline
\hline 
\multicolumn{8}{l}{Baseline}\tabularnewline
\hline 
 &  &  &  &  &  & & \tabularnewline
$Speculator \ Ratio$ & 0.033 & 0.081{*}{*} & 0.114{*}{*}&  0.191{*}{*} & 0.295{*}{*} & 0.378{*}{*} & 0.526{*}{*}\tabularnewline
 & (1.559) & (2.125) & (2.031) & (2.483) & (2.306) & (2.450) & (2.552)\tabularnewline
 &  &  &  &  &  & & \tabularnewline
R2 & 0.009 & 0.027 & 0.034 & 0.070 & 0.098 & 0.110 & 0.127\tabularnewline
 &  &  &  &  &  & & \tabularnewline
\hline 
\multicolumn{8}{l}{Month Fixed Effects}\tabularnewline
\hline 
 &  &  &  &  &  & & \tabularnewline
$Speculator \ Ratio$ & 0.063{*}{*} & 0.119{*}{*} & 0.166{*}{*}{*} & 0.228{*}{*} & 0.332{*}{*} & 0.421{*}{*} & 0.747{*}{*}{*}\tabularnewline
 & (2.042) & (2.183) & (2.196) & (2.353) & (2.208) & (2.324) & (2.660)\tabularnewline
 &  &  &  &  &  & & \tabularnewline
R2 & 0.091 & 0.103 & 0.127 & 0.141 & 0.144 & 0.127 & 0.181\tabularnewline
 &  &  &  &  &  & \tabularnewline
 \hline
\multicolumn{8}{c}{Panel B: Frequent Traders}\tabularnewline 
& $r_{t\to t+1}$ & $r_{t\to t+2}$ & $r_{t\to t+3}$ & $r_{t\to t+4}$ & $r_{t\to t+6}$ & $r_{t\to t+8}$ & $r_{t\to t+12}$\tabularnewline
\hline
\multicolumn{8}{l}{Baseline} \tabularnewline
\hline 
 &  &  &  &  &  & &  \tabularnewline
$Speculator \ Ratio^{low}$ & -0.025 & -0.003 & 0.014 & 0.002 & 0.065 & 0.079 & -0.045\tabularnewline
 & (-0.627) & (-0.071) & (0.247) & (0.022)& (0.660) & (0.580) & (-0.283)\tabularnewline
 &  &  &  &  &  & \tabularnewline
$Speculator \ Ratio^{high}$ & 0.050 & 0.082{*} & 0.105{*} & 0.188{*}{*}{*} & 0.253{*}{*} & 0.326{*}{*} & 0.556{*}{*}\tabularnewline
 & (1.405) & (1.760) & (1.846) & (2.714) & (2.208) & (2.274) & (2.571)\tabularnewline
 &  &  &  &  &  & & \tabularnewline
R2 & 0.012 & 0.027 & 0.035 &  0.070 & 0.100 & 0.113 & 0.130\tabularnewline
 &  &  &  &  &  & & \tabularnewline
\hline 
\multicolumn{8}{l}{Month Fixed Effects} \tabularnewline
\hline 
 &  &  &  &  &  & & \tabularnewline
$Speculator \ Ratio^{low}$ & -0.026 & -0.013 & 0.008 & -0.022 & 0.037 & 0.076 & 0.042\tabularnewline
 & (-0.652) & (-0.254) & (0.117) & (-0.275) & (0.345) & (0.551) & (0.231)\tabularnewline
 &  &  &  &  &  & & \tabularnewline
$Speculator \ Ratio^{high}$ & 0.078{*}{*} & 0.125{*}{*} & 0.160{*}{*} & 0.237{*}{*}{*} & 0.308{*}{*} & 0.375{*}{*} & 0.718{*}{*}{*}\tabularnewline
 & (2.105) & (2.381) & (2.385) & (2.794) & (2.280) & (2.271) & (2.878)\tabularnewline
 &  &  &  &  &  & & \tabularnewline
R2 & 0.094 & 0.104 & 0.127 & 0.142 & 0.145 & 0.130 & 0.183\tabularnewline
 &  &  &  &  &  & & \tabularnewline
\hline
\end{tabular}
\end{table}
{\footnotesize\par}

\subsection*{Frequent vs Infrequent Traders\label{subsec:Frequent-vs-Infrequent}}

Thus far, we find that emission allowance prices, on average, increase
subsequent to positive net purchases by firms, and decrease after
net sales. If these results are driven by asymmetric information,
with some operators exploiting private information to profit, we would
expect the return predictability power to be stronger for trades conducted
by firms that frequently transact in the emission allowance market.

To test this prediction, we calculate the number of trades companies engage in over the sample
period. We then create two subsamples based on the median number of
trades. For both subsamples, we calculate the variable ``$Speculator \ Ratio$''
as the net flow from person holding accounts to operator holding accounts
divided by the total transactions from person holding accounts
to operator holding accounts for each month. We denote $Speculator \ Ratio^{high}$
and $Speculator \ Ratio^{low}$ as the variable ``$Speculator \ Ratio$'' for the firms above
and below the sample median, respectively.

Panel B of Table \ref{tab:Predictability} presents the results from
time-series return predictability regressions using both $Speculator \ Ratio^{high}$
and $Speculator \ Ratio^{low}$. We first report  the baseline results without month
fixed effects and then also report the results with month fixed effects.
When both $Speculator \ Ratio^{high}$ and $Speculator \ Ratio^{low}$ are included in the regressions,
the coefficient estimates of $Speculator \ Ratio^{high}$ are always positive,
with or without month fixed effects. For the results without month
fixed effects, the coefficient estimates of $Speculator \ Ratio^{high}$ start
to be significant at the two-month horizon and remain significant
for longer horizons. For the results with month fixed effects, the
coefficient estimates of $Speculator \ Ratio^{high}$ are always significant at
the 5\% level or beyond. On the other hand, the coefficient estimates
of $Speculator \ Ratio^{low}$ are never significant in any of the specifications.
Overall, we find that the return predictability results are mainly
driven by emission allowance trading by frequent traders. This is
consistent with the notion that the excess trading of certain firms,
documented in Section \ref{sec:Trading-Patterns}, is driven by these
firms exploiting their private information about predictable price
trends of emission allowances. 

\section{Discussion}

The dramatic rise in carbon emissions poses a catastrophic threat through its contribution to climate change and global warming. Commonly considered approaches to controlling pollution are command-and-control policies~\citep{glaeser2001reason,jacobsen2023regulating}, carbon taxes~\citep{nordhaus1992optimal,aghion2016carbon}, and cap-and-trade policies~\citep{coase1960problem,montgomery1972markets,weitzman1974prices}. With the implementation of the Kyoto Protocol and the support of the US government, cap-and-trade systems have emerged as a preferred mechanism to curb the increase of carbon dioxide in the atmosphere~\citep{grubb2012cap}. Beyond the EU ETS, cap-and-trade systems are becoming more widely used in practice~\citep{newell2014carbon} and have been adopted in large economies such as California~\citep{fowlie2012emissions} and, more recently, China~\citep{liu2022challenges}.There has been recent interest in analyzing the implications of the Coase Theorem to environmental externalities~\citep{deryugina2021environmental,zaklan2023coase}. Our paper is the first to study the whole cap-and-trade carbon market as a financial market.

Empirical evidence indicates that cap-and-trade systems can achieve a reduction in emissions ~\citep{dobbeling2024systematic}. Our paper uses asset pricing methods and granular transaction data to uncover significant inefficiencies in the prominent cap-and-trade system which can substantially limit its effectiveness and are provide evidence of intense speculative trading by informed participants.



\pagebreak{}

\begin{singlespace}
\bibliographystyle{plainnat}
\bibliography{bib_emissions}
\end{singlespace}

\end{document}